\documentstyle[sprocl]{article}

\newcommand{\talpha}{\tilde{\alpha}}
\newcommand{\bbox}{\mathbf}
\newcommand{\tmu}{\tilde{\mu}}

\newcommand{\e}{{\rm e}}

\newcommand{\nn}{\nonumber \\}

\newcommand{\xp}{x^\prime}

\newcommand{\dy}{(dy)}

\newcommand{\dzi}{(dz_1)}
\newcommand{\dzii}{(dz_2)}

\newcommand{\dx}{(dx)}
\newcommand{\dxp}{(dx^{\prime})}

\newcommand{\dslash}{\gamma\partial}

\newcommand{\Aslash}{\gamma A}
\newcommand{\Bslash}{\gamma B}

\newcommand{\cq}{{{\mathsf{q}}_1\cdot{\mathsf{q}}_2}}
\newcommand{\wq}{{{\mathsf{q}}_1{\bbox{\times}}{\mathsf{q}}_2}}
\newfont{\fib}{cmfi10 at 10pt}
 
\newcommand{\pt}{\mbox{\fib{t}}}

\newcommand{\Tr}{{\rm Tr}}

\newcommand{\A}{{\cal A}}

\newcommand{\tD}{{\tilde{D}}}

\newcommand{\D}{{\cal D}}

\newcommand{\J}{{\cal J}}
\newcommand{\K}{{\cal K}}

\newcommand{\ie}{{\it i.e.}\ }

\newcommand{\fg}{\hspace{.1cm}^{\ast}F}
\newcommand{\jg}{\hskip .03cm ^{\ast}\hskip -.03cm j}
\newcommand{\Jg}{\hskip .03cm ^{\ast}\hskip -.03cm J}
\newcommand{\Gg}{\hspace{.1cm}^{\ast}G}

\def\be{\begin{equation}}
\def\ee{\end{equation}}
\def\bea{\begin{eqnarray}}
\def\eea{\end{eqnarray}}

\begin{document}
\begin{flushright}
OKHEP-00-05
\end{flushright}
\vskip .1cm
\title{EIKONAL SCATTERING OF MONOPOLES AND DYONS IN DUAL QED}

\author{LEONARD GAMBERG\footnote{Invited talk given at
``$5^{\rm th}$ Workshop on QCD'', Villefranche-sur-Mer, France, 
3-7 Jan. 2000.}${}^{,}$\footnote{E-mail: gamberg@mail.nhn.ou.edu}
and~KIMBALL A. MILTON\footnote{E-mail: milton@mail.nhn.ou.edu}}

\address{Department of Physics and Astronomy, University of Oklahoma\\
Norman, OK 73019}


\maketitle\abstracts{
The quantum field theory of electron-point
magnetic monopole interactions and dyon-dyon interactions,
based on the string-dependent ``nonlocal'' action of Dirac and Schwinger
is developed.
We demonstrate that a nonperturbative quantum
field theoretic formulation
can be constructed resulting in a string {\em independent}
cross section for monopole-electron and dyon-dyon scattering. 
Such calculations can be done only by using nonperturbative
approximations such as the eikonal and not by some
mutilation of lowest-order perturbation theory.}

\section{Introduction}\label{sec:intr}
%

The topic of magnetic charge has received enormous attention
since Dirac~\cite{dir} demonstrated its existence was
consistent with quantum mechanics provided the
quantization condition (in rationalized 
units) ${eg/\ 4\pi=N/\ 2}$ 
is satisfied. Here $e$ and $g$ are the strength of electric 
and magnetic charges,
respectively, and $N$ denotes an integer.
In the case of dyons, particles containing
both magnetic and electric charge,
the Schwinger generalization~\cite{sch1,sch2} 
\bea
\frac{e_{a}g_{b}-e_{b}g_{a}}{4\pi}=
\Bigg\{
\begin{array}{l}
\frac{N}{2}\, , {\mbox{unsymmetric}}\\
\mbox{\small $N$}\, , \mbox{symmetric}
\end{array}
\Bigg\}, 
\label{dyon}
\eea 
is invoked. (``Symmetric'' and ``unsymmetric'' refer to the presence or
absence of dual symmetry in the solutions of Maxwell's equations.)

With the advent of  non-Abelian theories, classical composite
monopole solutions were (theoretically)  
discovered.~\cite{non}  Their  mass 
would be of the order of the relevant gauge-symmetry breaking scale,
which for grand unified theories (GUT) is of order
$10^{16}$ GeV or higher.  However,
there are models where the electroweak symmetry
breaking can give rise to monopoles of mass $\sim 10$ TeV.~\cite{pres}
Yet, even the latter are not  accessible to accelerator experiments, 
so limits on heavy monopoles depend either on cosmological considerations,
or detection of cosmologically produced (relic) monopoles
impinging upon the earth or moon.~\cite{relic}
However, {\it a priori}, there is no reason that Dirac/Schwinger monopoles
or dyons of arbitrary mass might not exist.  In this respect, it 
is important to set limits below the 1 TeV scale in direct
 accelerator based experiments.~\footnote{Such an 
experiment is currently in progress at the University of Oklahoma,
\cite{kal} where we have set limits on {\it direct\/} monopole
production at Fermilab up to several hundred GeV.  This 
is an improvement over previous limits.~\cite{acc}  See 
also Ref.~\cite{gam2} for critique of theories underlying indirect searches. }

It is envisaged that if monopoles are sufficiently light, they would be
produced by a Drell-Yan type of process occurring in $p\overline p$ collisions
at the Tevatron.  The difficulty is to derive a reasonable estimate of the
elementary process $q\overline q\to\gamma^*\to M\overline M$, where $q$
stands for quark and $M$ for magnetic monopole.  
Attempts to incorporate monopoles consistently into
relativistic quantum field theory have met with mixed success.  
Weinberg, and soon thereafter Rabl,~\cite{wei}
demonstrated that the charge-monopole scattering amplitude, calculated in the
one-photon-exchange approximation, is a function of the
{\em Dirac string singularity}. Making matters worse, 
the value of the vertex coupling implied by 
$\alpha_{g}=g^2/{4\pi}\approx 34N^2$, calls 
into question any approach based on a badly divergent
perturbative expansion in $\alpha_{g}$.  
Although the early efforts using 
a Feynman-rule perturbation theory 
resulted in string-dependent cross--sections,
subsequently {\em ad hoc\/} 
assumptions were invoked to render the resulting cross-sections 
string independent (see Ref.~\cite{gam1} for details). 
In contrast, studying the {\it formal\/} behavior of Green's functions
in the relativistic quantum field theory of electrons and monopoles, both 
Schwinger~\cite{sch1,sch2} and Brandt {\it et al.}~\cite{zwa} 
demonstrated Lorentz-string and gauge invariance.  

However, with the exception of one instance~\cite{urr78}
such demonstrations have been conspicuously 
absent at the phenomenological level.\footnote{This is  surprising 
because one expects that the invariant  non-relativistic scattering result 
(see Ref.~\cite{mil} and references therein)
corresponds in a certain kinematic regime to a infinite summation of a 
particular {\em subclass} of Feynman diagrams.}
This deficiency  stems from the fact that in most 
phenomenological treatments of  
charge-monopole processes the ``string independence'' of 
the quantum field theory and the strength of the coupling are treated as 
separate issues. In fact, these two points are intimately 
related.  The lesson to be learned 
from the formal and non-relativistic demonstrations of Lorentz and 
string invariance is this:  Because the quantization condition 
is  intimately tied to the demonstration of 
Lorentz invariance, the latter can only be demonstrated using
a method which does not rely on perturbation 
theory~(see~\cite{gam1} for further discussion).

In view of the necessity of 
establishing a reliable estimate for monopole production in 
accelerators in order to be able to set bounds on monopole masses,  
it is important to put the theory 
of dual quantum electrodynamics (dual QED)
on a firmer foundation. With that in
mind we present our results.~\cite{gam1}

\section{Dual  Electrodynamics}

For a spin $1\over2$ monopole,
a minimal generalization of the QED action~\cite{sch2,gam1} for 
charge-monopole interactions 
expressed in terms of the vector potential $A_\mu$ 
and field strength tensor $F_{\mu\nu}$ (\ie, in a first-order formalism) is
\bea
W&=&\int\dx\bigg\{-\frac{1}{2}F^{\mu\nu}(x)
\left(\partial_\mu A_\nu\left(x\right)-\partial_\nu A_\mu\left(x\right)\right)+
\frac{1}{4}F_{\mu\nu}(x)F^{\mu\nu}(x)
\nn
&&
\hspace{-.75cm}+\, \bar\psi(x)\left(i\dslash+e{\Aslash}(x) 
-m_{\psi}\right)\psi(x)
+\bar\chi(x)
\left(i\dslash+g{\Bslash}(x)-m_{\chi}\right)\chi(x)\bigg\},
\label{act1}
\eea
where it is assumed that the electrically and magnetically charged
particles are spin $1/ 2$.
The resulting Maxwell's equations,
which imply the dual conservation of electric and magnetic
currents, $j_\mu$ and $\jg_\mu$,
necessitates  the introduction of the Dirac string singularity. 
The Dirac string function  satisfies the differential equation 
$\partial_\mu f^{\mu}(x)=\delta(x)$, which has the formal
solution $f^{\mu}(x)=n^\mu\left(n\cdot\partial\right)^{-1}\delta(x)$,
where $n^\mu$ is an arbitrary vector.\footnote{Here we 
have chosen the string to satisfy the oddness condition
(this is the ``symmetric'' solution) $f^\mu(x)=-f^\mu(-x)$,
corresponding to Schwinger's integer 
quantization condition.~\cite{mil}}
The field equations resulting from $\delta W=0$ are 
$\partial_{\nu}F^{\mu\nu}=j^{\mu}$ and
\be
F_{\mu\nu}=\partial_{\mu}A_{\nu}-\partial_{\nu}A_{\mu}+\Gg_{\mu\nu},
\label{fs}
\ee
where
\be
G_{\mu\nu}(x)=\int \dy \left(f_{\mu}(x-y)\jg_{\nu}(y)
-f_{\nu}(x-y)\jg_{\mu}(y)\right).
\label{gten}
\ee
The auxiliary dual field $B_\mu$ is defined as  a functional of field-strength
and depends on the string function,
\be
B_{\mu}(x)=-\int \dy f^\nu\left(x-y\right)\fg_{\mu\nu}(y).
\label{gfb}
\ee
Of course, the monopole field satisfies the Dirac equation
\be
\left(i\dslash +g\Bslash(x) - m_{g}\right)\chi(x)=0\, .
\label{diracm2}
\ee
From Eq.~(\ref{gfb}) we find that  $B_\mu$ satisfies
$\int (d x') f^{\mu}(x-x^{\prime})B_{\mu}(x^{\prime})=0$,
which is a special case of a gauge-fixed vector field 
defined in terms of the field strength through an {\em inversion}
formula, Eq.~(\ref{gfb}).
Similarly, we are at liberty to restrict the 
vector potential, $A_\mu$
to a hypersurface in field space embodied in the
inversion formula
\be
A_{\mu}(x)=-\int \dy f^\nu\left(x-y\right)\ F_{\mu\nu}(y)\, ,
\label{gfa}
\ee
which we denote as {\it string-gauge\/}, 
$\int (d x') f^{\mu}(x-x^{\prime})A_{\mu}(x^{\prime})=0$.
The photon kernel derived from the corresponding gauge fixed
action now possesses an inverse
\be
D_{\mu\nu}(x)=\Bigg[g_{\mu\nu}
-\frac{n_{\mu}\partial_{\nu}+n_{\nu}\partial_{\mu}}{(n\cdot\partial)}
+\left(1-\frac{1}{\kappa}\frac{(n\cdot\partial)^{2}\partial^{2}}{n^{2}}\right)
\frac{n^{2}\partial_{\mu}\partial_{\nu}}{(n\cdot\partial)^{2}}\Bigg]D_+(x)\, ,
\ee
where $D_+(x)$ is the massless scalar propagator,
\be
D_+(x)=\frac{1}{-\partial^{2}-i\epsilon}\delta(x)\, ,
\ee
which enables us to write 
an integral equation, expressing the vector potential, $A_\mu$ 
($B_\mu$) in terms of the electric (magnetic) and magnetic (electric)
currents.
We generalize these classical integral equations to one
point Green functions in obtaining
the generating function for Green's functions in dual QED.

\section{Quantization of Dual QED: Schwinger-Dyson Equations}

Using a path integral formulation to quantize the
string-dependent action is by no means straightforward.
In order to unambiguously develop the generating functional 
for physical processes we make use of Schwinger's quantum 
action principle,~\cite{sch51} 
where we write the vacuum persistence 
amplitude for Green functions in the presence of external sources,
$Z({\cal J})=\langle 0_+\left|\right. 0_-\rangle^{\cal J}$
for the charge-monopole system.
That is, under an arbitrary  variation, 
\be
\delta\langle 0_+\left|\right. 0_-\rangle^{\J}=
i\langle 0_+\left| \delta W(\J)\right| 0_-\rangle^{\J},
\ee
where $W(\J)$ is the action given in Eq.~(\ref{act1}) externally 
driven by the sources, $\J$, which for the present case are
given by the set of terms 
\be
W({\cal J})=W+\int (dx)\left\{J^\mu A_\mu+{}^* J^\mu B_\mu+\bar\eta\psi
+\bar\psi\eta+\bar\xi\chi+\bar\chi\xi\right\}.
\ee
Given the  one-point functions (${\cal O_{\mu}}$ is the field conjugate
to the source $\J^{\mu}$)
\be
\frac{\delta\log Z(\J)}{i\delta \J^\mu(x)}=
\frac{\langle 0_+|{\cal O}_\mu(x)| 0_-\rangle^\J}
{\langle 0_+\left|\right. 0_-\rangle^{\cal J}},
\label{onept}
\ee
we solve the corresponding coupled Schwinger-Dyson equations for the vacuum amplitude~\cite{gam1}, subject to the gauge conditions
\be
\int (dx')f^\nu(x-x')
{\delta\langle 0_+|0_-\rangle_0^{\cal J}\over\delta J^\nu(x')}
=0,\quad
\int(dx')f^\mu(x-x'){\delta\langle 0_+|0_-\rangle^{
\cal J}\over\delta {}^*J_\mu(x')}=0.
\label{gctwo}
\ee
Since any expansion in $\alpha_g$ or $eg$
is not practically useful we recast the solution into a functional 
form better suited for a nonperturbative calculation of the four-point
Green's function.~\cite{gam1}  
For dyons, the different species of which are labeled by the index $a$ this is
\bea
Z(\J)&=&
\exp\Bigg\{\frac{i}{2}
\int(dx)(dx')\K^\mu(x)\D_{\mu\nu}\left(x-x^\prime\right)
\K^\nu\left(x^\prime\right)\Bigg\}
\nn
&&\times
\exp\Bigg\{\frac{i}{2}\int(dx)(dx')
\frac{\delta}{\delta\bar{\A}_{\mu}(x)}
\D_{\mu\nu}(x-x')\frac{\delta}{\delta\bar{\A}_{\nu}(x')}\Bigg\}
\nn
&&\times
\exp\bigg\{ i\sum_a\int(dx)(dx')\bar{\zeta}_a\left(x\right)
G_a(x,\xp|\bar{\A_a})\zeta_a\left(\xp\right)\bigg\}
\nn
&&\times 
\exp\bigg\{-\sum_a\int_{0}^{1} dq \,
\Tr\gamma\bar{\A_a} G_a(x,x|q\bar{\A_a})\bigg\},
\label{masterd}
\eea
where $\A_a=e_aA+g_aB$, $\zeta_a$ is the source for the dyon of species $a$,
and a matrix notation is adopted,
\bea
\K^{\mu}(x)&=&
\left(
\begin{array}{c}
J(x)\\ \Jg(x)
\end{array}
\right),
\hspace{.5cm}
\frac{\delta}{\delta\bar{\A}_{\mu}(x)}=
\left(
\begin{array}{c} 
\delta/\delta\bar{A}_{\mu}(x) \\
\delta/\delta\bar{B}_{\mu}(x)
\end{array} 
\right),
\nn
\nn
\hspace{-3cm}\D_{\mu\nu}\left(x-x^\prime\right)&=&
\left(
\begin{array}{cc}
D_{\mu\nu}\left(x-x^\prime\right) & -\tD_{\mu\nu}\left(x-x^\prime\right) \\
\tD_{\mu\nu}\left(x-x^\prime\right) & D_{\mu\nu}\left(x-x^\prime\right)
\end{array}
\right)\, .
\eea
We use the shorthand notation for the ``dual
propagator'' that couples magnetic to electric charge
\be
\tD_{\mu\nu}\left(x-x'\right)=\epsilon_{\mu\nu\sigma\tau}
\int (dx'')
D_+\left(x-x''\right)
\partial^{\prime\prime\sigma} f^\tau\left(x''-x'\right).
\label{dualp}
\ee
The two-point fermion Green's functions 
in the background of the stationary
photon field $\bar{A},\bar{B}$ are given by
\be
G(x,x'|\bar{\A})=\langle x|
(\gamma p +m-{\bar{\A}})^{-1}| x'\rangle.
\ee

\section{Eikonal Approximation for Dyon-Dyon and 
Charge-Monopole Scattering}

To calculate the dyon-dyon scattering cross section
we obtain the four--point Green's
function for this process from Eq. (\ref{masterd}),
\be
G\left(x_1,y_1;x_2,y_2\right)=
\frac{\delta}{i\delta\bar{\zeta}_1(x_1)}
\frac{\delta}{i\delta{\zeta}_1(y_1)}
\frac{\delta}{i\delta\bar{\zeta}_2(x_2)}
\frac{\delta}{i\delta{\zeta}_2(y_2)}
Z(\J)\bigg|_{{\cal J}=0}.
\label{gnfnct3}
\ee
The subscripts on the sources refer to the two different dyons.
As a first step in analyzing the string dependence
of the scattering amplitudes,
we study high-energy forward scattering
processes where {\em soft} photon exchanges
dominate.  Diagrammatically,
in this kinematic regime we restrict attention to 
that subclass in which there are no closed fermion loops
and the photons are exchanged 
between fermions.~\cite{nac91}
This amounts to quenched-ladder
approximation 
where the linkage operators, $\L$,
connect two fermion propagators 
via photon exchange.  We can read this off from
Eq.~(\ref{masterd}):
\be
\e^{\L_{12}} =\exp\Bigg\{i
\int \dx\dxp\frac{\delta}{\delta\bar{\A}^{\mu}_1(x)}
\D^{\mu\nu}\left(x-\xp\right)
\frac{\delta}{\delta \bar{\A}^{\nu}_2(\xp)}\Bigg\},
\label{link}
\ee
so Eq.~(\ref{gnfnct3}) takes the form
\bea
G\left(x_1,y_1;x_2,y_2\right)&=&
-\e^{\L_{12}}
G_1(x_1,y_1|\bar\A_1)G_2(x_2,y_2|\bar\A_2)\Big|_{\bar{A}=\bar{B}=0},
\label{scatt}
\eea
where we express the two-point function using the proper-time 
parameter representation of an ordered exponential.
The soft, nonperturbative effects of
the interaction between electric and magnetic charges
dominate in the region where the momentum exchanged by the photons 
is small compared to the center of mass energy
$s=-(p_1+p_2)^2$,   \ie $t/ s\ll\, 1$. This amounts to 
the Bloch-Nordsieck~\cite{blo37} 
or {\it eikonal approximation}.~\cite{fri65,aba} 
This approximation substantially simplifies evaluating
the path-ordered exponentials in Eq.~(\ref{scatt}).
They are now exponentials of linear functionals of the gauge field.

\subsection{High Energy Scattering Cross Section}
\label{s:ea}

Using the identity,
\be
\e^{\L}=1+\int_0^1 da\, \e^{a\L}\L
\label{par}
\ee
one obtains to the following form of the four-point
Green function, Eqs.~(\ref{scatt}),
\bea
\hspace{-.25cm}G(x_{1},y_{1};x_{2},y_{2})&=&
\int_0^1 da\int \dzi\dzii\overline{\cal D}_{\mu\nu}(z_1-z_2)\e^{a\L_{12}}
\nn
&& 
\hspace{-2.5cm}\times
G_1(x_{1},z_1|\bar{\A}_1)\gamma_\mu
G_1(z_1,y_{1}|\bar{\A}_1)
G_2(x_{2},z_2|\bar{\A}_2)\gamma^\nu
G_2(z_2,y_{2}|\bar{\A}_2)\Bigg|_{\bar A=\bar B=0},
\label{mxptl}
\eea 
where, $\overline{\cal D}_{\mu\nu}(x)$ represents the combination
of propagators,
\be
\overline{\cal D}_{\mu\nu}(z_1-z_2)=
\cq D_{\mu\nu}(z_1-z_2)-\wq\tD_{\mu\nu}(z_1-z_2),
\ee
and the charge combinations invariant under duality transformations 
are 
\be
\cq=~e_1e_2~+~g_1g_2\quad {\mbox{\rm and}}\quad \wq=e_1g_2-g_1e_2\, \, .
\ee
Choosing a space like string, $n^\mu=(0,\bbox{n})$
and the incoming momenta to be in the $z$
direction, in the center of momentum frame,
the M{\o}ller amplitude, $M(s,t)$ is given by
\bea
\hspace{-.5cm}
M(s,t)&=&\frac{-i}{2\pi}\int_0^1 da\int d^2{\bbox{x}}\, \e^{i\bbox{q}\cdot\bbox{x}}
\bar{u}(p^{\prime}_{1})\gamma^{\mu}u(p_{1})
\bar{u}(p^{\prime}_{2})\gamma^{\nu}u(p_{2})\, 
\e^{ia\Phi_{n}\left(s,t ; x\right)}
\nn
&&
\hspace{-1.25cm}\bigg\{
g_{\mu\nu}\cq K_0\left(\mu\left|{\bbox{x}}\right|\right)
-\epsilon_{\mu\nu\sigma\tau}\wq
n^\tau\frac{\partial}{\partial n_\sigma}
{1\over2}\int\,\frac{d\pt}{\pt}
K_0
\left(\mu\left|\left(\bbox{x}+\pt\bbox{n}\right)\right|\right)\bigg\},
\label{smp5}
\eea
where in this kinematic regime, the eikonal phase 
is ($\mu$ is the photon mass; ${\tilde{\mu}}=\e^{\gamma}\mu/ 2$ and 
$\gamma$ is Euler's constant)
\be
\Phi_{n}\left(s,t;x\right)=\frac{1}{2\pi}\left\{\cq
\ln\left(\tilde{\mu}\left|{\bbox{x}}\right|\right)
-\wq\arctan\left[\frac{\bbox{\hat n}\cdot\bbox{x}}
{\bbox{\hat z}\cdot\left(\bbox{\hat n}\times\bbox{x}\right)}\right]
\right\}.
\ee
First, we calculate the helicity amplitudes 
in the high-energy limit, $p^0\gg m$. 
In performing the integral over the impact parameter 
care must be taken since the arctangent function is discontinuous
when the $xy$ component of $\bbox{\hat n}$
and $\bbox{x}$ lie in the same direction.  
However, requiring that the eikonal phase factor $e^{i\Phi_n}$ 
be continuous, necessarily leads to the 
Schwinger quantization condition (\ref{dyon}):
$\wq= 4N\pi$.
Now using the integral form for 
the Bessel function of order $\nu$
\be
i^\nu\/J_\nu(t)=\int_0^{2\pi}\frac{d\phi}{2\pi}
\e^{i\left(t\cos\phi-\nu\phi\right)}\, ,
\ee
we find the dyon-dyon  scattering amplitude to be
\be
M(s,t)=\frac{ s}{M_1M_2}{2\pi\over q^2}(N-i\talpha)
\e^{i2N\psi}
\left(\frac{4\tmu^2}{q^2}\right)^{i\talpha}
\frac{\Gamma\left(1+N+i\talpha\right)}
{\Gamma\left(1+N-i\talpha\right)}\, ,
\ee
where $\tilde{\alpha}=\cq/4\pi$, and $\psi$ is the angle between
$\bbox{q}$ and $\bbox{n}$.

This result is almost identical in structure to the non-relativistic
form of the scattering amplitude for the Coulomb potential, which result
is recovered by setting $N=0$ (see, for example, Ref.~\cite{aba}.)
Following the standard convention  we calculate
the spin-averaged cross section for dyon-dyon scattering
in the high energy limit,
\be
\frac{d\sigma}{dt}=
\frac{\left(\cq\right)^2+\left(\wq\right)^2}{4\pi t^2}.
\label{dyoncross}
\ee
For the case of charge-monopole scattering $e_1=g_2=0$,
this result, coincides with that
found by Urrutia~\footnote{Utilizing Schwinger's functional 
source theory \cite{sch66b}  
in the context of a zeroth order eikonal approximation
Urrutia \cite{urr78}  demonstrated string independence of the 
charge-monopole scattering cross section, although in 
his treatment the currents are approximated by those of
classical point particles.} 
which  is  also  string independent as a consequence 
of the quantization condition.  

\section{Corrections}

To go beyond the regime of soft or infrared photon exchange 
requires a detailed  analysis of factorization 
of soft and hard contributions and their correlations. 
Such effects have been widely studied in the context of 
phenomenologically based hadronic interacting
field theories~\cite{gas71} and more recently in the
context of diffractive scattering~\cite{nac91} in QCD
and in the world-line formalism.~\cite{kor94,gr98}

As a reasonable first step we impose corrections on the eikonal amplitude
by relaxing the ``high energy'' approximation on the spinors 
in Eq.~(\ref{smp5}) while assuming that 
the soft contributions are dominated by the 
eikonal phase.~\cite{fri} This result, which we expect 
to be a better and better approximation
in the high energy limit (for a given $t$), \ie $t/ s \rightarrow 0$,
obeys the expected scaling behavior, for electron monopole scattering
\bea
s^2\frac{d\sigma}{dt}=\frac{(eg)^2}{4\pi}
\frac{1}{t^2}\left(s+t\right)^2\Rightarrow f\left(\frac{t}{s}\right).
\eea
Assuming that we have extended the kinematic range of this result
beyond the low $t^2$ limit, we can consider using
the analytic properties of the scattering amplitude to 
calculate the Drell-Yan production amplitude in the $t$-channel.
Detail will be presented in future publication.

\section{Conclusions}

We  have given a complete formulation, in modern quantum
field theoretic language, of an interacting electron-monopole and dyon-dyon
systems. The challenge is to apply these equations to the calculation 
of monopole and dyon
processes.  Perturbation theory is useless, not only because of the strength
of the coupling, but also because the graphs are fatally string-
(or gauge-) dependent.  The most obvious nonperturbative technique for 
transcending these limitations in scattering processes lies in the 
high energy regime where the 
eikonal approximation is applicable; in that limit, our formalism 
generalizes  an early  lowest-order result of Urrutia and
charts the way to include systematic corrections.  
More problematic is the treatment of monopole production 
processes.
We will apply the techniques and results found here to the Drell-Yan
production of monopole-antimonopole processes, and obtain phenomenologically
relevant estimates for the accelerator production of monopole-antimonopole
pairs.

In addition we have also detailed
how the Dirac string dependence disappears from physical quantities.
It is by no means a result of string 
averaging or a result of dropping
string-dependent terms (see Ref.~\cite{gam1} for details);
but rather, a result of summing the soft contributions to 
the dyon-dyon or electron-monopole process.  
At the level of the eikonal
approximation and its corrections
one might suspect the occurrence of a 
factorization of hard string-independent
and soft string-dependent contributions in a manner similar to that
argued recently in strong-coupling QCD.
There is  reason
to believe that inclusion of soft-hard correlations
in the scattering will
not spoil this consistency. 

\section*{Acknowledgments}
L.G. would like to thank Herb Fried and the  
organizers of ``Fifth Workshop on QCD'' for
the invitation to present this work. 
This work is supported in part by the 
Department of Energy.

\end{document}